\documentstyle[12pt]{article} 
 
\def\be{\begin{equation}}
\def\ee{\end{equation}}
\def\a{\alpha}
\def\g{\gamma}
\def\d{\delta}
\def\e{\epsilon}
\def\s{\sigma}

\def\c{\tilde{c}}
\def\b{\tilde{b}}
\def\n{\{n\}}

\def\S{\tilde{S}} 
 
\def\F{\hat{F}}
\def\Re{\mbox{Re}}
\def\l{\lambda} 

\def\pd{\partial}
\def\vac{|0\rangle} 

\def\ra{\rangle}
\def\la{\langle}
\def\ln{\mbox{ln}}
\def\sin{\mbox{sin}}
\def\ctg{\mbox{ctg}}
\def\tg{\mbox{tg}}

\begin{document}

\begin{center} 
{\large   Exactly solvable discrete BCS - type Hamiltonians \\
               and the Six-Vertex model.}
\end{center}

\begin{center}
{\large  A.A. Ovchinnikov  }
\end{center}

\begin{center}
{\it Institute for Nuclear Research, RAS, Moscow, 117312, Russia} 
\end{center}

\begin{center}
{\it (September, 2002)}
\end{center}

\vspace{0.1in}

\begin{abstract}

 We propose the new family of the exactly solvable discrete state BCS - 
type Hamiltonians based on its relationship to the six-vertex model 
in the quasiclassical limit both in the rational and the trigonometric cases. 
We establish the relation of the BCS Hamiltonian and its eigenfunctions to 
the form of the monodromy matrix in the F-basis. 
Using the Algebraic Bethe Ansatz method
for the standard BCS model with equal coupling the expression 
for the general scalar product and the determinant expressions for the 
physically interesting correlation functions for the finite number of sites 
which can be used in the numerical and analytical computations are obtained. 
We also compare the correlators with the results obtained by means of the 
variational method.

\end{abstract}

\vspace{0.4in}

              {\bf 1. Introduction.}

\vspace{0.2in}

  At present time the exactly solvable discrete-state Bardeen, Cooper and 
Schrieffer (BCS) model for the superconductivity \cite{BCS} attracts much 
attention in connection with the problems in different areas of physics such 
as superconductivity, nuclear physics, physics of ultrasmall metallic grains 
and color superconductivity in QCD.  
   The exact solution of the discrete- state BCS model is especially important 
for the study of the superconducting correlations in the atomic nuclei and the 
ultrasmall mettalic grains since due to the finite number of particles the 
description in terms of the grand canonical ensemble \cite{BCS} 
(in contrast to the microcanonical ensemble) is obviously not correct. 
    For all of the above mentioned problems and also for the many-body 
problems of fermions with the long-range interaction, it is desirable to find 
the integrable BCS- type Hamiltonians with the attraction of the Cooper pairs, 
depending on the momentum (on the indices of sites in the discrete - state 
model) and on the occupation numbers on the other sites.   
    Note that at present time the study of both the continuum and the discrete 
BCS model in the case of the equal spacing of the energy levels is not 
completed since the possibility of varying the number of pairs with the 
filling of the part of the energy levels by the single- electron states 
was not considered.  
    Another important problem which motivates the study of different 
modifications of the BCS Hamiltonian is to find the realistic integrable BCS - 
type models which apart from the interaction of pairs, include the 
interactions describing the disintegration of pairs i.e. the hopping of  
two single electrons (fermions) to another energy levels independently of 
whether they form a Cooper pair or not. In this case the determinant 
expressions for the correlators obtained in the present paper 
could be useful as well.  
    The studying of the excitations of different kind in the case of the 
microcanonical ensemble for the BCS model is also an interesting problem. 
    For the case of the continuum limit of the BCS model as well as for the 
case of the microcanonical ensemble the analytical and the numerical 
calculations for various correlation functions are important.

The eigenvectors and the eigenvalues of the discrete BCS Hamiltonian where 
first constructed by Richardson \cite{RS},\cite{R65} in the context of nuclear 
 physics. Later the model was applied also to the case of the Bose gas 
\cite{R68}. The norm and the simplest correlation functions have been studied 
in ref. \cite{R65}, \cite{R68}. 
Recently the integrability of the BCS model was proved by Cambiaggio, Rivas 
and Saraceno \cite{CRS}. The set of the commuting operators which have much in 
common with the Gaudin magnets \cite{G} was constructed. The connection of the 
Gaudin magnets with the six- vertex model was pointed out by Sklyanin 
\cite{Sklyan}. The continuum limit of the BCS model for the equally spaced 
distribution of the energy levels was first considered by Gaudin \cite{GAUDIN} 
and later by Richardson \cite{R77}, who also performed the numerical 
calculations \cite{R77}, \cite{RPR}, and was found in agreement with the 
original (variational) BCS treatment \cite{BCS}. Recently the solution of the 
model based on the off-shell Algebraic Bethe Ansatz construction \cite{BF} was 
given by Amico, Falci and Fazio \cite{AFF}. 
The numerical calculation of the correlation functions for the finite systems  
was presented by Amico and Osterloh \cite{AO} using the method of calculation 
of the scalar products, based on the generating function, proposed by Sklyanin 
\cite{SKL}. Nevertheless the calculations are quite involved and restricted to 
the systems with the small number of sites $N$. Therefore the explicit 
determinant expressions for the correlation functions of the model are highly 
desirable. Recently the connection of the BCS model with the twisted 
inhomogeneous six-vertex model was elaborated in ref's \cite{DP} and 
\cite{Zhou}, where the possibility of computation of the correlators with the 
help of the Algebraic Bethe Ansatz method was pointed out. 
However, the suitable determinant formulas for the correlators 
have not been obtained.  The new multiparameter families of the BCS -type 
models connected with the trigonometric six-vertex model where proposed in 
ref's \cite{ADO}, \cite{Richardson} and studied in more detail in 
ref. \cite{ALMO}. The connection of the BCS model with the conformal field 
theory and the modified Knizhnik-Zamolodchikov equations \cite{KZ} have been 
studied by Sierra (for example, see \cite{Sierra}).  
The discrete BCS Hamiltonian was generalized to the case of arbitrary 
degeneracy of different energy levels of the Cooper pairs \cite{G} 
and to the case of the Dicke model \cite{D} (for example, see \cite{J}).

The main goal of the present paper is the calculation of the correlators in 
the BCS - type models with the help of the methods developed in the context 
of the Algebraic Bethe Ansatz approach to the six-vertex model. 
We present the new determinant expressions for the basic correlation 
functions of the BCS model. We also review and propose the new approaches 
to the construction of various integrable BCS - type Hamiltonians.  

 In the first part of the paper we briefy reveiw some of the recent results 
on the BCS -type models, in particular, the derivation of the generalized  
BCS models from the six-vertex model both in the rational and the 
trigonometric cases in a way which is similar to that of ref's \cite{DP} - 
\cite{Richardson}. We present the new way of construction 
of the BCS - type models with the interaction depending on the lattice 
sites by means of considering the limit of the transfer matrix spectral 
parameter $t\to\infty$. The several examples of the Hamiltonians with the 
position - dependent interaction are  presented explicitly.  
We establish the relation of the BCS Hamiltonian and its eigenfunctions 
to the form of the monodromy matrix in the F-basis which also leads to 
the generalization of the BCS model to the case of the interaction between 
pairs depending on the occupation numbers on the other sites. 
Let us stress, that the formalism of the F- basis is used not to obtain the 
known formulas for the scalar products for the six-vertex model, but to 
develop the new formalism for the construction of various BCS- like integrable 
models. 

  In the second part of the paper we present the results on the calculation  
of various correlation functions for the BCS model using the Algebraic Bethe 
Ansatz method for the six-vertex model. First, we obtain the simple 
expressions for the scalar products and the formfactors of the model taking 
the quasiclassical limit in the corresponding expressions for the six-vertex 
model. Then, to obtain the expressions for the correlators, we use the 
commutational relations for the operators directly in the quasiclassical limit 
in order to reduce the problem to the calculation of the scalar products.    
We obtain the new determinant expressions for the different correlation 
functions which are usuful for the numerical and the analytical calculation 
of the correlators. 
Let us stress, that our results for the correlators are different from the 
results obtained previously in ref's \cite{AO}, \cite{Zhou}, and allow one  
the much more simple numerical evaluation of the correlators, since in all  
cases the correlators are represented as the determinants or a finite sum 
of the determinants.

The content of the paper is as follows. 
In Section 2 we propose the new family of the exactly solvable 
discrete BCS - type Hamiltonians based on its relationship to the six-vertex 
model in the quasiclassical limit both in the rational and the 
trigonometric cases. 
We present the examples of the Hamiltonians with the interaction between 
the pairs depending on the energy levels.     
We also review the results \cite{ADO} on the BCS - type integrable models  
with the double set of parameters. 
In Section 3 we establish the relation of the 
BCS Hamiltonian and its eigenfunctions to the form of the 
monodromy matrix in the F-basis which also leads to 
the generalization of the BCS model to the case of the interaction between 
pairs depending on the occupation numbers on the other sites. 
The expression for the general scalar product and   
the determinant expressions for the physically interesting correlation 
functions for the finite number of sites which can be used in numerical 
computations are obtained in Section 4. 
The comparison with the expressions for the correlation functions obtained 
with the help of variation on the parameters is presented in the 
Appendix B. 
Thus we show that no special technique of the type proposed in 
ref.\cite{SKL} for the calculation of correlators 
for the BCS model and for the Gaudin magnets is required. 
The results for the correlation functions for the Gaudin magnets are also 
interesting from the theoretical point of view.  
The results obtained can be useful for studying the correlation 
functions for the XXZ quantum spin chain. 
For completeness some of the results on the diagonalization of the BCS 
Hamiltonian are also presented in the Appendix A. 
As an additional application of the solution of the BCS model we present 
the solution of the modified Knizhnik-Zamolodchikov equations 
\cite{KZ} for the correlators of the conformal field theory 
($SU(2)$ WZW- model) in the Appendix C.  
We present in the Appendix D the brief review of the Gaudin's solution 
of the BCS model in the thermodynamic limit.

\vspace{0.4in}

{\bf 2. BCS Hamiltonian and the Six-Vertex model.} 

\vspace{0.2in}

The BCS Hamiltonian with an arbitrary parameters $\e_i$ has the form: 
\be
H=\sum_{i=1}^N \e_i n_i - gS^{+}S^{-},
\label{bcs}
\ee
where $n_i=b_i^{+}b_i$ is the number of the hard-core bosons 
(electron pairs), $S^{+}=\sum_ib_i^{+}$, $S^{-}=\sum_ib_i$ and the coupling 
constant $g$ is positive, which corresponds to the attraction.   

To solve the Schrodinger equation for the Hamiltonian (\ref{bcs}) and its 
generalizations (see below), instead of the usual transfer matrix of the 
six-vertex model with twisted boundary conditions, we consider the following 
transfer matrix for the twisted inhomogeneous six-vertex model  
\be
Z(t,\{\xi\})=\mbox{Tr}_0
\left( \S_{10}(\xi_1,t)\S_{20}(\xi_2,t)\ldots \S_{N0}(\xi_N,t)\right), 
\label{z}
\ee
which can be equivalently represented as the trace of the monodromy matrix 
in the auxiliary space $0$,  
\[
Z(t,\{\xi\})=\mbox{Tr}_0\left(T_0(t)\right),~~~~~
T_0(t)=\left( 
\begin{array}{cc}
A(t) & B(t) \\
C(t) & D(t) 
\end{array} \right)_0, 
\]
where $\xi_i$ are the inhomogeneity parameters. 
The matrices $\S_{i0}(\xi_i,t)$ are equal to 
\[
\S_{i0}(\xi_i,t)=K_0 S_{i0}(\xi_i,t) 
\]
where $K_0$ is the twist matrix and 
$S_{i0}$ is the usual S-matrix of the six-vertex model obeying the Yang-Baxter 
equation $S_{12}S_{13}S_{23}=S_{23}S_{13}S_{12}$:
\[
K_0= \left( 
\begin{array}{cc}
e^{\eta/2gN}   & 0 \\
0 &  e^{-\eta/2gN} 
\end{array} \right)_0,~~~~
S_{12}(t_1,t_2)=t_1-t_2+\frac{\eta}{2}\left(\sigma_{1}\sigma_{2}\right)
\]
(we denote $(\sigma_{1}\sigma_{2})=\sum_a\sigma_{1}^a\sigma_{2}^a,~a=x,y,z$)
for the rational case. 
Due to the well known property of the matrix $R_{00'}=P_{00'}S_{00'}$
($P_{12}=\frac{1}{2}(1+(\sigma_1\sigma_2))$ - is the permutation operator) 
$\left[K_{0}K_{0'};R_{00'}\right]=0$
(it does not matter if the standard twist angle is imaginary or real) 
the matrices $\S_{i0}(\xi_i,t)$ obey the usual Yang-Baxter equation  
$R_{00'}\S_{i0}\S_{i0'}=\S_{i0'}\S_{i0} R_{00'}$ and the transfer 
matrices (\ref{z}) commute at different values of the spectral parameters 
$\left[ Z(t);Z(t')\right]=0$. 
The Algebraic Bethe Ansatz method (for example, see \cite{FST}) can be readily 
applied to the monodromy matrix defined in (\ref{z}) to obtain the spectrum of 
the BCS model and evidently leads to the same results as for 
the transfer matrix with the usual twisted boundary conditions 
$Z(t)=\mbox{Tr}_0(K_0T_0(t))$, where $T_0(t)$ is the usual monodromy matrix 
constructed from the matrices $S_{i0}(\xi_i-t)$ and 
$K_0$ is the diagonal matrix corresponding to the 
total twist angle $\eta/2g$. However the construction presented above allows 
for another generalization which will be considered later. 
Considering the limit $\eta\to0$ we get up to a factor $(\xi_i-t)$: 
\be 
\S_{i0}=1+(\eta/2gN)\sigma_0^z+(\eta/2(\xi_i-t))(\sigma_0\sigma_i)+O(\eta^2), 
\label{slimit}
\ee
and retaining the terms of order $O(\eta^2)$, 
we obtain the following Hamiltonian depending on the parameter $t$: 
\be
H(t)=-\frac{1}{2g}\sum_{i=1}^N\frac{1}{(t-\xi_i)}\sigma^z_i + \frac{1}{2}
\sum_{i<j}\frac{1}{(t-\xi_i)(t-\xi_j)}\left(\sigma_i\sigma_j\right),
\label{ht}
\ee
where $\sigma^z_i$ can be substituted by the number of the hard-core bosons 
$\sigma^z_i=2n_i-1$. 
Since the operators $H(t)$ commute at different values of the parameter $t$, 
one can obtain the set of the commuting operators, which  generalize the 
Gaudin magnets \cite{G}, taking the limit $t\to\xi_i$: 
\be
H_i=-\frac{1}{g}n_i+
\frac{1}{2}\sum_{l\neq i}\frac{(\sigma_i\sigma_l)}{(\xi_i-\xi_l)}, 
~~~~~\left[H_i;H_j\right]=0~~~\left[H;H_i\right]=0.  
\label{hi}
\ee
Note that since the commuting operators can be defined up to an 
additive constant, we choose the operators (\ref{hi}), omitting the 
constant term in the operator $\sigma^z_i=2n_i-1$. 
Note also that in these equations the operator $(\sigma_i\sigma_j)$ 
can be represented through the hard-core boson operators $b_i^{+},b_i$ as 
\[
(\sigma_i\sigma_j)=2(b_i^{+}b_j+ b_j^{+}b_i)+(2n_i-1)(2n_j-1).
\]
These operators also commute with the Hamiltonian (\ref{bcs}) which was 
first found in ref.\cite{CRS}. In fact, considering the limit $t\to\infty$ 
in eq.(\ref{ht}) and retaining the terms of order $\sim 1/t^2$, we obtain 
exactly the Hamiltonian (\ref{bcs}) with $\e_i=\xi_i$ (in what follows 
we omit the normalization factors and the additive constants depending on 
the total number of the hard - core bosons).   
Alternatively one can consider the following linear combination of the 
operators $H_i$ to obtain the same Hamiltonian: 
\be
H= -g~\sum_{i=1}^{N}\xi_i H_i=\sum_{i=1}^N\xi_i n_i - gS^{+}S^{-},
\label{Richard}
\ee
which coincides with the expression (\ref{bcs}) up to an additive constant, 
depending on the number of bosons $M$. 
The following correspondence between the spin and the 
hard-core boson operators is used: $S^{\pm}=\sum_{i}S_i^{\pm}$, 
$S_i^a=\frac{1}{2}\sigma_i^a$, 
$S_i^{\pm}=b_i^{+}(b_i)=\frac{1}{2}(\sigma_i^x\pm i\sigma_i^y)$.  
One can also consider an arbitrary linear combinations of the operators $H_i$ 
with the coefficients $\e_i\neq\xi_i$. Then the following generalization of 
the BCS Hamiltonian appears: 
\be
H=\sum_{i=1}^N\e_i n_i-\frac{g}{2}\sum_{i<j}\frac{(\e_i-\e_j)}{(\xi_i-\xi_j)}
\left(\sigma_i\sigma_j\right).
\label{xh}
\ee
Let us note also that using the different notations the Hamiltonian 
(\ref{ht}) can be represented in the form: 
\be
H=\frac{1}{g}\sum_{i=1}^N \e_i n_i + 
\sum_{i<j} \e_i \e_j \left(\sigma_i\sigma_j\right). 
\label{new}
\ee
Thus, in general, one can construct the Hamiltonians with the interaction 
between the Cooper pairs depending on the momentum of the pairs. 
Another Hamiltonian of this type can be obtained by considering the 
derivative of the transfer matrix over the spectral parameter $t$. 
In this way the Hamiltonian takes the form similar to (\ref{new}) with 
an extra factor $\e_i^{-1/2}+\e_j^{-1/2}$ contained in the sum of the 
second term. In general, since $H(t)$ eq.(\ref{ht}) depends on the 
additional parameter $t$ and commute at different $t$, 
$\left[H(t);H(t')\right]=0$, one can take the 
derivatives over $t$ and consider the linear combination of the 
operators $H^{(n)}(t)$ at arbitrary $t$ ($H=\sum_{n}C_{n}H^{(n)}(t)$).  
Thus, we obtain the new Hamiltonians of the BCS type (\ref{new}) 
with the infinite number of additional parameters $C_n$.   
The other way to obtain new Hamiltonians is to consider the 
decomposition of $H(t)$ in the powers of $1/t^n$ at large $t$. 
In this way the new models with the dependence on the additional set of 
parameters can be obtained. 
Clearly, the similar procedures can be applied for the case of the 
trigonometric (hyperbolic) six-vertex model. 

 To obtain the eigenvalues of the BCS Hamiltonian one can start with the 
well known procedure of diagonalization of the transfer matrix (\ref{z})- 
the Algebraic Bethe Ansatz method (for example, see \cite{FST}, \cite{Vega}). 
The eigenstates are represented as
\be
|\phi(t_i)\ra=B(t_1)B(t_2)\ldots B(t_M)\vac, 
\label{bbb}
\ee
where the parameters $t_1\ldots t_M$ obey the system of Bethe Ansatz equations 
which do not depend on the spectral parameter $t$: 
\be
e^{-\eta/g}\prod_{\a=1}^{N}
\left(\frac{t_i-\xi_{\a}-\eta/2}{t_i-\xi_{\a}+\eta/2}\right)= 
\prod_{\a=1}^{M}\left(\frac{t_i-t_{\a}-\eta}{t_i-t_{\a}+\eta}\right). 
\label{ba}
\ee
Decomposing eq.(\ref{ba}) to the first order in the small parameter $\eta$, 
we obtain the Richardson's equations 
\be
\sum_{\a=1}^{N}\frac{1}{t_i-\xi_{\a}}-2~\sum_{\a\neq i}^{M}
\frac{1}{t_i-t_{\a}}= - \frac{1}{g}.
\label{rich}
\ee
The corresponding eigenvalue for the Hamiltonian (\ref{bcs}) is obtained 
as the limit at $t\to\infty$ of the eigenvalue of the transfer matrix $Z(t)$: 
\[
E=\sum_{i=1}^M~t_i.
\]
The eigenvalues of the conserved operators $H_i$ (\ref{hi}) are easily 
evaluated from the eigenvalue of the transfer matrix (for the rational case)   
\[
\Lambda(t)=e^{-\eta/2g}\prod_{\a}\frac{\xi_{\a}-t-\eta/2}{\xi_{\a}-t}
\prod_{i}\frac{t_i-t+\eta}{t_i-t}+e^{\eta/2g}\prod_{\a}
\frac{\xi_{\a}-t+\eta/2}{\xi_{\a}-t}\prod_{i}\frac{t-t_i+\eta}{t-t_i},
\]
which in the limit $\eta\to0$ up to the terms of order $\eta^2$ 
and $t\to\xi_i$ produces the eigenvalues of the operators $H_i$ (\ref{hi}): 
\be
E_i=\frac{1}{2}\sum_{\a\neq i}\frac{1}{(\xi_i-\xi_{\a})}+
\sum_{j}\frac{1}{(t_j-\xi_i)} 
\label{ei}
\ee
(here we take into account the definition of the constant term 
corresponding to the operators $H_i$ (\ref{hi})). 
In the quasiclassical limit the operators $B(t)$, $C(t)$ reduce to the 
operators $\Sigma^{\pm}(t)$ obeying the commutational relations which can be 
easily obtained from the basic commutational relations for the elements 
of the monodromy matrix (Yang-Baxter equation): 
\be
B(t)\to\Sigma^{+}(t),~~ C(t)\to\Sigma^{-}(t),~~~~  
\Sigma^{\pm,z}(t)=\sum_{i=1}^{N}\frac{\sigma_i^{\pm,z}}{t-\xi_i},
\label{sigma}
\ee
\[
\left[\Sigma^{-}(t);\Sigma^{+}(t')\right]=
 2 \frac{\Sigma^{z}(t)-\Sigma^{z}(t')}{t-t'},~~~~~
\left[\Sigma^{z}(t);\Sigma^{\pm}(t')\right]=
\mp \frac{\Sigma^{\pm}(t)-\Sigma^{\pm}(t')}{t-t'} 
\]
Note that one could built up the eigenstates of the Hamiltonian (\ref{bcs}) 
directly in terms of the operators $\Sigma^{\pm}(t)$ \cite{AFF} 
(see also the Appendix A).  
Thus we see that instead of the off-shell Bethe Ansatz approach \cite{BF} used 
in ref.\cite{AFF} one can use the usual on-shell Bethe Ansatz equations for 
the six-vertex model.

  Let us consider the trigonometric S- matrix. Repeating the steps 
leading to the Hamiltonian (\ref{ht}) in the rational case 
(i.e. first taking the limit $\eta\to 0$) we obtain in the 
trigonometric case the following Hamiltonian: 
\be
H(t)=-\frac{1}{2g}\sum_{i=1}^N\ctg(t-\xi_i)n_i+
\sum_{i<j}\frac{1}{\sin(t-\xi_i)\sin(t-\xi_j)} \hat{Y}_{ij}+ 
\sum_{i<j}\ctg(t-\xi_i)\ctg(t-\xi_j) \hat{A}_{ij},
\label{htsin}
\ee
where the following notations are used: 
\[
\hat{Y}_{ij}=b_i^{+}b_j+b_j^{+}b_i,~~~\hat{A}_{ij}=n_i n_j +(1-n_i)(1-n_j).
\]
That is the one - parameter generalization of the Hamiltonian (\ref{ht}).
First, one can proceed in the following way. Taking the limit $t\to\xi_i$, 
one readily obtains the trigonometric analogs of the commuting operators 
$H_i$ (\ref{hi}) 
\be
H_i=-\frac{1}{2g}n_i+\sum_{l\neq i}\left(\frac{1}{\sin(\xi_i-\xi_l)} 
\hat{Y}_{il}+\frac{1}{\tg(\xi_i-\xi_l)}\hat{A}_{il}\right), 
\label{hisin}
\ee
and considering the linear combination 
$\sum_{i}\e_{i}H_i$, we get the Hamiltonian 
\be
H=-\frac{1}{2g}\sum_{i=1}^{N}\e_{i}n_{i}+
\sum_{i\neq j}\frac{\e_i-\e_j}{\sin(\xi_i-\xi_j)}\hat{Y}_{ij}+
\sum_{i\neq j}\frac{\e_i-\e_j}{\tg(\xi_i-\xi_j)} \hat{A}_{ij}.
\label{trig}
\ee
The second possibility is to consider the limit $t\to\infty$ in the hyperbolic 
version of eq.(\ref{htsin}). Then we get the following integrable Hamiltonian 
($\g_i=e^{\xi_i}$): 
\[
H=-\frac{1}{2g}\sum_{i=1}^{N}\g_{i}^2 n_{i}+
\sum_{i\neq j} \g_i \g_j \hat{Y}_{ij}.
\]
The omitted terms vanish in the limit considered.

Using the different notations the last Hamiltonian can be represented  
in the form: 
\be
H=-\frac{1}{2g}\sum_{i=1}^{N}\e_i n_{i}+
\sum_{i\neq j}\sqrt{\e_i}\sqrt{\e_j}\left( b_i^{+}b_j+b_j^{+}b_i \right).
\label{trigon}
\ee
Note that the last expression is different from the Hamiltonian (\ref{new}) 
obtained in the case of the rational six-vertex model. 
As in the rational case the equation (\ref{htsin}) is more general 
than eq.(\ref{trig}) and taking the derivatives over $t$ one also can 
obtain the various generalizations of the Hamiltonian (\ref{trigon}). 
For the hyperbolic case which is obtained by considering $\xi_i$ 
as an imaginary parameters the Hamiltonian (\ref{trig}) in the limit 
$g\to\infty$ can be considered as the Hamiltonian of the open spin chain 
with the long-range interaction.  
Let us mention also that the other possibility to construct the exactly - 
solvable BCS- like Hamiltonians is to consider the various bilinear 
combinations of the form $H=\sum_{i,j}f_{ij}H_{i}H_{j}$ with different 
coefficients $f_{ij}$. 

Let us comment on the form of the eigenstates and the equations for the 
eigenvalues for the trigonometric case. The analogs of the operators 
$\Sigma^{\pm}(t)$ in this case are 
\[
\Sigma^{\pm}(t)=\sum_{i}\frac{1}{\sin(t-\xi_{i})}\sigma_{i}^{\pm},  
\]
and the analogs of Richardson's equations in the trigonometric case, 
which can be obtained from the trigonometric version of the equations 
(\ref{ba}), are  
\[
\sum_{\a=1}^{N}\frac{1}{\tg(t_i-\xi_{\a})}-2\sum_{\a\neq i}^{M} 
\frac{1}{\tg(t_i-t_{\a})}=-\frac{1}{g}, 
\]
which represent the conditions for the common eigenvectors for the operators 
(\ref{hisin}) and the Hamiltonian depending on the double set of parameters 
(\ref{trigon}). Considering the quasiclassical limit of the eigenvalue 
$\Lambda(t)$, we obtain, omitting the factor 
$\prod_{\a}\sin(\xi_{\a}-t)$, the expression 
\[
\Lambda(t)=e^{-\eta/2g}\prod_{i=1}^M\frac{\sin(t_i-t+\eta)}{\sin(t_i-t)} + 
e^{\eta/2g}\prod_{i=1}^M\frac{\sin(t_i-t-\eta)}{\sin(t_i-t)} 
\prod_{\a=1}^N\frac{\sin(\xi_{\a}-t+\eta)}{\sin(\xi_{\a}-t)}, 
\]
where the definition of the parameter $t$ differs from the definition in the 
rational case by the shift $t\to t+\eta/2$ and considering the limit 
$t\to\xi_i$, we get 
\[
E_i=1/2g - \sum_j\ctg(t_j-\xi_i). 
\]
Here the difference with the expressions (\ref{ei}) is due to the 
different definitions of the operators $H_i$ (note that 
$P_{ij}=\hat{A}_{ij}+\hat{Y}_{ij}=(1/2)(1+(\sigma_i\sigma_j))$).

Let us mention also the generalization of the construction proposed above. 
Consider the monodromy matrix constructed from the S- matrices 
$\S_{i0}(\xi_i,t)=K_0^{(i)}S_{i0}(\xi_i,t)$ where the twist matrix 
$K_0^{(i)}$ now depends on the site $i$: 
\[
K_0^{(i)}= \left(
\begin{array}{cc}
e^{\eta c_i/2gN} & 0 \\
0 & e^{-\eta c_i/2gN} 
\end{array}
\right)_0, 
\]
where $c_i$ are an arbitrary parameters. Clearly in this case the 
Algebraic Bethe Ansatz method can be used in the usual way. 
Let us denote $c_0=\sum_{i}c_i$. Then, first taking the limit $\eta\to0$,  
we will obtain the same Hamiltonian with the coupling constant depending 
on the constant $c_0$ only.  However, considering this limit in the case  
of $c_0=1$ and the parameters $c_i$ are of order 
$c_i\sim N/\eta$, then, to obtain the Hamiltonian of the type (\ref{bcs}), 
one should commute all the matrices $K_0^{(i)}$ to the left, which 
effectively leads to the gauge-like transformation for the operators 
$b_i^{+}$, $b_i$, which enter the operators $S^{\pm}$ in the 
Hamiltonian (\ref{bcs}).

 The discrete BCS model (\ref{bcs}) can be generalized to the case of 
arbitrary degeneracy of different energy levels corresponding to the 
energies of the Cooper pairs which is equivalent to the case of an 
arbitrary spin $s_i$ assigned to the site $i$ with the energy $\e_i$. 
Previously, in Section 2, the case $s_i=1/2$ was considered. 
One can show that the limit $s_i\to\infty$ for the number of 
sites $\e_i$ corresponds to the special generalization of the  
Dicke model. Thus, using the general six-vertex model, the eigenfunctions 
and the eigenvalues of the Hamiltonian can be obtained. 
Clearly, instead of the elementary $S$-matrix one can take the 
following Lax operator in the equation (\ref{z}): 
\[
L_{i0}(\xi_i-t)=\xi_i-t+\eta\left(\sigma S_i\right),
\]
where $S_i^a,~a=x,y,z$ are the spin operators with the value of the spin 
 $s_i$, $(S_i)^2=s_i(s_i+1)$. Considering the quasiclassical limit of the 
transfer matrix (\ref{z}) we obtain instead of (\ref{bcs}) the Hamiltonian 
\be
H=\sum_{i=1}^N \e_i S^z_i - gS^{+}S^{-},
\label{siham}
\ee
where $S^{\pm}=\sum_{i=1}^{N}S^{\pm}_i$. Clearly, in terms of the 
initial BCS model the integer parameters $2s_i$ correspond to the 
degeneracies $2s_i$ of $i$-th level with the energy $\e_i$. 
The construction of the eigenstates is the same as in Section 2 and
the equations (\ref{rich}) become  
\be
\sum_{\a=1}^{N}\frac{2s_i}{t_i-\xi_{\a}}-2~\sum_{\a\neq i}^{M}
\frac{1}{t_i-t_{\a}}= - \frac{1}{g}, 
\label{sirich}
\ee
while the eigenvalues of the Hamiltonian have the same form   
$E=\sum_{i=1}^N~t_i$. The eigenstates can be constructed 
also with the help of the Gaudin operators  
$\Sigma^{\pm}(t)=\sum_{i=1}^{N}\frac{S_i^{\pm}}{t-\xi_i}$ with
$\xi_i=\e_i$. For the BCS model with the equal degeneracies $2s$ 
at each site the equations (\ref{sirich}) take the form
\be
(2s)\sum_{\a=1}^{N}\frac{1}{t_i-\xi_{\a}}-2\sum_{\a\neq i}^{M}
\frac{1}{t_i-t_{\a}}= -\frac{1}{g}, 
\label{srich}
\ee
Considering the limit $s\to\infty$ of the spin at the single site, 
and using the expressions of  the spin operators at this site 
through the Holstein- Primakoff bosons, 
$\left[\phi;\phi^+\right]=1$, $S^z=\phi^+\phi-s$, 
$S^+=\phi^+(2s-\phi^+\phi)^{1/2}$, $S^-=(2s-\phi^+\phi)^{1/2}\phi$,   
we get after rescaling the parameters the Dicke Hamiltonian \cite{D}: 
\be
H=\omega\phi^+\phi+\sum_{i=1}^{N}\e_i n_i-g\left(S^+\phi+S^-\phi^+\right).  
\label{bd}
\ee
Thus the spectrum and the eigenstates of this Hamiltonian can also be 
obtained using the appropriate limiting procedure of that of the transfer 
matrix of the six-vertex model.  Remarkably, 
this model can be generalized to the case of several species of the 
oscillators $\phi_i^+,\phi_i$ with different frequencies $\omega_i$.

\vspace{0.4in}

{\bf 3. Connection with the monodromy matrix in the F- basis.}

\vspace{0.2in}

Let us show how the above results can be obtained using the operator 
expression of the monodromy matrix in the $F$ -basis, the basis obtained
with the help of the factorizing operator $F$ introduced in ref.\cite{MS}.   
One can construct the operator $F = F_{1...N}$ which diagonalizes the 
operator $A(t)$ \cite{MS},\cite{KMT},\cite{O} ($A^F(t)=F^{-1}A(t)F$). 
Using the notations 
\[
\c(t)=\frac{\phi(t)}{\phi(t+\eta)},~~~  
\b(t)=\frac{\phi(\eta)}{\phi(t+\eta)}, 
\]
where $\phi(t)=t$ for the 
rational case and $\phi(t)=\sin(t)$ for the trigonometric case,  
the diagonal operator $A^F(t)$ has the following form: 
\be
A^F(t)=\prod_{i=1}^N\left(\c(\xi_i-t)(1-n_i)+n_i\right).
\label{af}
\ee
Let us briefly mention some of the properties of the operator $F$. 
The explicit form of the operator $F$ is 
\be
F_{12\ldots N}=\F_1\F_2\ldots \F_N,~~~~\F_i = (1-\hat{n}_i)+T_i \hat{n}_i,  
\label{f}
\ee
where $\hat{n}_i$ is the operator of the number of particles (spin up) at the 
site $i$ and the operator $T_n$ is given by the equation 
\[
T_n = S_{n+1,n}S_{n+2,n}\ldots S_{Nn}. 
\]
One can obtain the following formulas for the matrix elements of 
the operator $F$ \cite{O} in the following form: 
\[
F_{\{m\}\{n\}}=\la\{m\}|B(\xi_{n_1})B(\xi_{n_2})\ldots B(\xi_{n_M})\vac, 
\] 
where the sets of coordinates $\{m\}$ and $\{n\}$ label the positions of the 
occupied sites. The similar expression can be obtained for the inverse 
operator $F^{-1}$. Apart from diagonalizing the operator $A(t)$, the operator 
$F$ is the factorizing operator \cite{MS} in the following sense. 
For any permutation of indices $\sigma\in S_N$ 
($S_N$ - is the group of permutations) we have the equation 
$F=F^{\sigma}R^{\s}$, where $F^{\sigma}_{12..N}=F_{\s1\s2..\s N}$ 
(including the permutation of the inhomogeneity parameters $\xi_i$) 
and $R^{\s}_{1...N}$ is the operator constructed from the $S$- matrices 
defined in such a way that for the permutation of the monodromy matrix 
$T_0^{\s}=T_{0,\s1\s2..\s N}$ we have $T_0^{\s}= (R^{\s})^{-1}T_{0}R^{\s}$. 
For the particular permutation $\sigma(\n)$ such that 
$\s 1=n_1,\ldots \s M=n_M$ ($n_1<n_2< \ldots <n_M$)  
the factorization condition is represented as 
$F(F^{\sigma(\n)})^{-1}=T_{n_1}..T_{n_M}$.     
To prove the factorizing property of the operator (\ref{f}) 
it is sufficient to consider only one particular permutation, for example, 
the permutation $(i,i+1)$, since all the other can be obtained as a 
superposition of these ones for different $i$. One can show, that  
$F=S_{i+1,i}F^{(i,i+1)}$, which evidently proves the factorization property.  

The matrix elements of the operators $B(t)$ and $C(t)$  
in the F - basis: $B^{F}(t)=F^{-1}B(t)F$ (and the same for $C(t)$) 
have the following form
\be
B^{F}(t)=\sum_{i}\sigma_{i}^{\dagger}~\b(\xi_{i}-t)\prod_{k\neq i} 
\left(\c(\xi_k-t)(\c(\xi_k-\xi_{i}))^{-1}(1-n_k)+n_k \right).  
\label{bf}
\ee
\be
C^{F}(t)=\sum_{i}\sigma_i^{-}~\b(\xi_i-t)\prod_{k\neq i} 
\left( \c(\xi_{k}-t)(1-n_k)+(\c(\xi_i-\xi_{k}))^{-1} n_k\right).  
\label{cf}
\ee
The operators (\ref{bf}) and (\ref{cf}) are quasilocal i.e. they describe  
flipping of the spin on a single site with the amplitude depending on the 
positions of the up-spins on the other sites of the chain. It is easily 
seen that in the limit $\eta\to0$ these operators reduce to the operators 
$\Sigma^{\pm}(t)$ (\ref{sigma}). The operator 
$D^F(t)$ can be found, for example, using the quantum determinant 
relation and has a (quasi)bilocal form. In fact, the following operator 
identity for the elements of the monodromy matrix (\ref{z}) can be derived: 
\[
D(t)A(t-\eta)-B(t)C(t-\eta)=
\prod_{\a}\left((t-\xi_{\a}+\eta/2)(t-\xi_{\a}-\eta/2)\right),  
\]
which is readily transformed to the F- basis. From this relation the explicit 
form of the operator $D^F(t)$ can be obtained. However perhaps 
the simplest way to obtain $D^F(t)$ is to use the well known 
basic commutational relation following from the Yang-Baxter equation: 
\[
\left[B(t);C(q)\right]=\frac{\b(q-t)}{\c(q-t)}
\left(D(t)A(q)-D(q)A(t)\right).
\]
Considering this equation in the F- basis in the limit $q\to\infty$, 
we obtain the following expression for the operator $D^F(t)$: 
\[
D^F(t)=A^F(t)+\left[B^F(t);C^{-}\right], 
\]
where the operators $A^F$, $B^F$ are given in (\ref{af}), (\ref{bf}) 
and the operator $C^{-}=\lim_{q\to\infty}(q/\eta)C^F(q)$ equals: 
\[
C^{-}=-\sum_{i}b_i\prod_{k\neq i}
\left((1-n_k)+(\c(\xi_i-\xi_k))^{-1}n_k\right). 
\]
To find the physically interesting BCS-type models it is not necessary 
to consider the quasiclassical limit. For example, considering the limit 
$t\to\infty$ for $Z^F(t)$ we obtain (omitting an overall factor $\eta/t$ 
and an additive constant depending on $M$) the Hamiltonian of the form 
\[
H=\left[B^{+};C^{-}\right], 
\]
where the operator $B^{+}$ equals 
\[
B^{+}=-\sum_{i}b_i^{+}\prod_{k\neq i}
\left(n_k+(\c(\xi_k-\xi_i))^{-1}(1-n_k)\right), 
\]
and the operator $C^{-}$ was defined above. 
This Hamiltonian has a (quasi)bilocal form and, apart from the set of the 
parameters $\{\xi\}$, depends on the additional parameter $\eta$. 

So the transfer matrix in the F-basis $Z^F(t)=A^F(t)+D^F(t)$
represents the Hamiltonian of the BCS -type model with the varying 
interaction of Cooper pairs (depending on the occupation numbers 
on the other sites) even without taking the quasiclassical limit. 
It is easily seen that in the limit $\eta\to0$ the results of the previous 
section are reproduced in such a way that 
the second term in eq.(\ref{ht}) corresponds to the term $D^F(t)$.    
Note also that since the pseudovacuum state is invariant with respect to the 
action of the factorizing operator $F\vac=\vac$, the eigenfunctions 
of $Z^F(t)$ have the form $\prod_{i}B^F(t_i)\vac$. As it was mentioned above, 
in the quasiclassical limit the operators $B^F(t)$ and $C^F(t)$ 
eqs.(\ref{bf}), (\ref{cf}) reduce to the operators $\Sigma^{+}(t)$ 
and $\Sigma^{-}(t)$ defined in eq.(\ref{sigma}).  
Concluding this section, let us stress once more that the Hamiltonian 
$Z^F(t)$, generalized by including the twist angle, contains the terms 
of the form $\sum_{i}\e_{i}n_i$ and the terms of the bilocal form 
$\sim b_i^{+}b_j$ with the amplitude depending on the occupation numbers 
on the other sites, and reduces to the BCS Hamiltonian (\ref{bcs}) 
in the quasiclassical limit with the corresponding twist angle.

\vspace{0.4in}

{\bf 4. Correlation functions.}

\vspace{0.2in}

Here we derive the analytical expressions for the simplest 
physically interesting correlation functions: 
$\la0|n_i\vac$, $\la0|b_i^{+}b_j\vac$, $\la0|S^{+}S^{-}\vac$ and 
$\la0|(\sigma_i\sigma_j)\vac$.  For simplicity we restrict ourselves 
to the rational case although the similar formulas can be easily 
obtained for the general trigonometric (hyperbolic) models.  
First, we consider the following scalar product:
\be
S_M(\{\l\},\{t\})=\la 0|C(\l_1)C(\l_2)...C(\l_M)
                    B(t_1)B(t_2)...B(t_M)\vac ,  
\label{scalar}
\ee 
where $\{\l\}$ and $\{t\}$ are the two sets of parameters, 
the set $\{t\}$ satisfies the Bethe Ansatz equations and 
$\{\l\}$ is an arbitrary set of parameters.  
According to the connection with the six-vertex model revealed in Section 2,  
to obtain the expression for the scalar product in the case when the set 
of the parameters $\{t\}$ satisfies the equations (\ref{rich}) one can use 
as a first step the known formula for the six - vertex model in the 
case when the parameters $\{t\}$ satisfy the usual Bethe Ansatz equations 
(\ref{ba}) \cite{O}, \cite{K}, \cite{S}, \cite{GMW}  
(see also the direct proof in ref. \cite{KMT} 
and in a different way in ref.\cite{O1}), 
and then decompose it in powers of $\eta$ 
(extract the leading power $\eta^{2M}$). The formulas for the XXX- spin 
chain are 
\be
S_M(\{\l\},\{t\})=\frac{1}{\prod_{i<j}(t_i-t_j)\prod_{j<i}(\l_i-\l_j)}
\det_{ij}(M_{ij}(t,\l)),
\label{m}
\ee
where the matrix $M_{ij}$ equals:  
\be
M_{ij}(t,\l) = \frac{\eta}{(t_i-\l_j)} 
 \left( \frac{a(\l_j)f^{+}(\l_j)}{t_{i}-\l_j +\eta} - 
        \frac{f^{-}(\l_j)}{t_{i}-\l_j -\eta} \right),
\label{mbethe} 
\ee
where the following notations are used:
\[
f^{\pm}(\l)=\prod_{\a=1}^{M}\left(t_{\a}-\l\pm\eta\right),
\]
and $a(\l)=\prod_{\a}\c(\xi_{\a}-\l)$. 
Note that in eq.(\ref{scalar}) the operators $B(t)$, $C(t)$ are normalized
in such a way that their quasiclassical limit is given by the operators 
(\ref{sigma}). 
From the equation (\ref{mbethe}) taking the limit $\l_i\to t_i$ one can 
easily obtain the formula for the norm of the Bethe eigenvector:  
\[
N_M(t)=\eta^M\frac{\prod_{i\neq j}(t_i-t_j+\eta)}{\prod_{i\neq j}(t_i-t_j)}
\det_{ij}\left[N_{ij}(t)\right], 
\]
where the matrix $N_{ij}$ can be represented in the form: 
\[
N_{ij}=\frac{2\eta}{(t_{ij}+\eta)(t_{ij}-\eta)},~~(i\neq j),~~~~~ 
N_{ii}= - \frac{\partial}{\partial t_i} \ln\left( a(t_i)) \right) -  
  \sum_{\a\neq i} \frac{2\eta}{(t_{\a i}+\eta)(t_{\a i}-\eta)}, 
\]
where $t_{ij}=t_i-t_j$ and $a(t)=e^{2\eta/g}\prod_{\a}\c(\xi_{\a}-t)$, 
$a(t)\to1+2\eta/g+\eta\sum_{\a}(t-\xi_{\a})^{-1}$ at $\eta\to0$.    
Extracting from the last expression the term of order $\eta^{2M}$, we 
get the norm of the eigenstate: 
\[
\la\phi(t)|\phi(t)\ra=\det_{ij}\left[N_{ij}(t)\right], 
\]
(here $\la\phi(t)|\phi(t)\ra=S_M(t,t)$) where the matrix $N_{ij}$ is given by 
\be
N_{ii}=\sum_{\a}\frac{1}{(t_i-\xi_{\a})^2}-\sum_{\a\neq i}
\frac{2}{(t_{i\a})^2},
~~~~N_{ij}=\frac{2}{(t_{ij})^2}, ~~~i\neq j. 
\label{norm}
\ee
The formula for the norm was first derived by Richardson \cite{R65}. 
Note that the norm does not depend on $\e_i$ explicitly. 
Similarly the general scalar product (\ref{scalar}) is given by the 
formula (\ref{m}) with the matrix $M_{ij}$ of the following form: 
\be
M_{ij}(\l,t)=\frac{1}{(t_i-\l_j)^2}\left(\prod_{\a}(t_{\a}-\l_j)
\right)\left[
\sum_{\a}\frac{1}{(\l_j-\xi_{\a})}-
2\sum_{k\neq i}\frac{1}{(\l_j-t_k)}+\frac{1}{g} \right]. 
\label{mrich}
\ee
Thus the scalar product (\ref{scalar}) with one Bethe eigenstate is obtained 
in the form  given by the equations (\ref{m}), (\ref{mrich}). Equivalently, 
since the product in the matrix (\ref{mrich}) depends only on the number 
of the column $j$ this scalar product can be also represented in the form 
\[
S_M(\{\l\},\{t\})=
\frac{\prod_{i,j}(t_i-\l_j)}{\prod_{i<j}(t_i-t_j)\prod_{j<i}(\l_i-\l_j)}
\det_{ij}(\hat{M}_{ij}(\l,t)),
\]
where the matrix $\hat{M}_{ij}(\l,t)$ is given by 
\[
\hat{M}_{ij}(\l,t)=\frac{1}{(t_i-\l_j)^2}
\left[\sum_{\a}\frac{1}{(\l_j-\xi_{\a})}-
2\sum_{k\neq i}\frac{1}{(\l_j-t_k)}+\frac{1}{g} \right]. 
\]
However for the calculation of the correlators in order to consider the 
limit $\l_i\to t_i$ for some of $\l_i$, it is more convenient to use the 
formulas (\ref{m}), (\ref{mrich}). 
The formula for the norm can be obtained directly by taking the limit 
$\l_i\to t_i$ in this expression for $S_M(\l,t)$ for the BCS model. 
Since the set $\{t\}$ satisfies the Richardson Bethe Ansatz equations 
(\ref{rich}) we obtain from (\ref{mrich}) in this limit the matrix elements 
\[
M_{ii}\to\prod_{\a\neq i}(t_{\a}-t_i) \left(
\sum_{\a}\frac{1}{(t_i-\xi_{\a})^2}-2\sum_{\a\neq i}
\frac{1}{(t_{i\a})^2}\right),~~~
M_{ij}\to\prod_{\a\neq j}(t_{\a}-t_j)\frac{1}{(t_i-t_j)^2},~~~i\neq j.
\]
Again, since the product which depend only on the number of the column $j$ 
can be written in front of determinant, taking into account the products in 
front of the determinant (\ref{m}), we obtain exactly the norm (\ref{norm}). 
For completeness, let us present another equivalent expression for the 
scalar product with one Bethe eigenstate which can also be used for the 
derivation of the norm and can be useful for taking different limits in the 
process of the calculation:    
\[
S_M(\l,t)=
\frac{\prod_{i\neq j}(t_i-\l_j)}{\prod_{i<j}(t_i-t_j)\prod_{j<i}(\l_i-\l_j)} 
\det_{ij}\left(\tilde{M}_{ij}(\l,t)\right),
\]
where the new matrix $\tilde{M}_{ij}$ is equal to 
\[
\tilde{M}_{ij}(\l,t)=\frac{(t_j-\l_j)}{(t_i-\l_j)^2}\left(
\sum_{\a}\frac{1}{\l_j-\xi_{\a}}-2\sum_{k\neq i}\frac{1}{\l_j-t_k} 
+\frac{1}{g}\right),  
\]
where the first term in the numerator also depends only on the index $j$. 

Now one can calculate the physically interesting correlation functions using 
the Algebraic Bethe Ansatz method. 
Although the expressions for some of the correlators can be obtained 
by the method of variations over the parameters \cite{R65}, \cite{G},  
which we present in the Appendix B, the determinant expressions obtained 
with the help of the Algebraic Bethe Ansatz method are different in the 
form (although equivalent to the variational ones). At the same time 
the direct Bethe Ansatz method allows for the computations of some of the 
correlation functions (such as $\la b_i^{+}b_j\ra$, for example)  
that are not accessible by the variational method and can be useful for 
the computation of the correlators in the different BCS- like models. 

Let us proceed with the calculation of the simplest correlation function 
$\la n_i\ra$ along the lines of ref.\cite{KMT} and using the 
technique developed above.  
First, we consider the action of the operator $A(\xi_i)$ at the state 
$|\phi(\l)\ra$ (\ref{bbb}) where $\l_i$ are not necessarily satisfy the 
Bethe Ansatz equations. Due to the symmetry of the problem, one can consider 
the site $\xi_1$.  One can use the formulas for the scalar product with 
Bethe eigenstate (\ref{m}), (\ref{mbethe}) taking subsequently the 
quasiclassical limit.  
Note that due to the symmetry of the Hamiltonian (\ref{bcs}) here 
it is not necessary to use the general formulae of the Quantum Inverse 
Scattering method for the six-vertex model \cite{KMTI}. 
However, the most simple way is to represent the 
eigenstate directly in terms of the operators (\ref{sigma}), 
\be
|\phi(t)\ra=\Sigma^{+}(t_1)\Sigma^{+}(t_2)\ldots\Sigma^{+}(t_M)\vac
\label{rstate}
\ee
and use the formula for the scalar product (\ref{mrich}). To calculate the 
expectation value $\la\phi(t)|n_1|\phi(t)\ra$ we act by the operator $n_1$ 
on the state (\ref{rstate}) using the following relation:  
\[
n_1\Sigma^{+}(t)=\Sigma^{+}(t)n_1+\frac{1}{(t-\xi_1)}b_1^{+},~~~n_1\vac=0.  
\]
The result can be represented in the form which allows one to apply the 
expression (\ref{mrich}) for the scalar product: 
\[
n_1|\phi(t)\ra=\sum_i\frac{1}{(t_i-\xi_1)}\lim_{\zeta\to\xi_1}(\zeta-\xi_1)
\left(\Sigma^{+}(t_1)\ldots\Sigma^{+}(\zeta)\ldots\Sigma^{+}(t_M)\right)\vac,  
\]
where the operator $\Sigma^{+}(\zeta)$ replace the operator 
$\Sigma^{+}(t_i)$ in eq.(\ref{rstate}). 
The factor $(\zeta-\xi_1)\to0$ in this formula is cancelled by the 
corresponding term in the denominator in the first sum in the expression 
(\ref{mrich}) for the matrix $M_{ij}$ at $\l_i=\zeta$. 
Next, one can use the following theorem. 
Consider the determinant of the sum of the two matrices 
$\det_{ij}(M_{ij}+a_{ij})$, where the second matrix $a_{ij}=c_j\phi_i$ -
is the matrix of rank 1. Then we have: 
\be
\det_{ij}\left(M_{ij}+a_{ij}\right) = \det_{ij}(M_{ij}) + 
\sum_{k=1}^{M} c_k \det_{ij}(M_{ij}^{(k)}), 
\label{cc}
\ee
where the matrices $M_{ij}^{(k)}$ differ from the initial matrix $M_{ij}$ 
only by the substitution of its $k$-th column by $\phi_i$:
\[
M_{ij}^{(k)}=(1-\delta_{jk})M_{ij}+ \delta_{jk}\phi_i. 
\]
Applying this theorem to the case of the average $\la\phi(\l)|n_1|\phi(t)\ra$, 
the determinant in eq.(\ref{mrich}) transforms into the sum of the 
determinants in (\ref{cc}) with 
\[
\phi_i=M_{ik}^{(k)}=\frac{1}{(t_i-\xi_1)^2}\prod_{\a}(t_{\a}-\xi_1),~~~
c_k=\prod_{\a\neq k}\left(\frac{\l_{\a}-\l_k}{\l_{\a}-\xi_1}\right)
\frac{1}{(\l_k-\xi_1)}. 
\]
Thus, we get the expression 
\[
\la\phi(\l)|n_1|\phi(t)\ra=\frac{1}{\prod_{i<j}(t_i-t_j)\prod_{j<i}(\l_i-\l_j)}
\left(\det_{ij}(M_{ij}+H_{ij})-\det_{ij}(M_{ij})\right), 
\]
where the matrix $H_{ij}=c_j\phi_i$ in the limit $\l_i\to t_i$ equals 
\[
H_{ij}(\l\to t)=\prod_{\a\neq j}(t_{\a}-t_j)\frac{1}{(t_i-\xi_1)^2}, 
\]
and the matrix $M_{ij}$ is given by the equation (\ref{mrich}). 
Taking the limit 
$\l_i\to t_i$ in the whole expression, we finally obtain the formula 
\be
\la\phi(t)|n_1|\phi(t)\ra=\det(N_{ij}+H_{ij})-\det(N_{ij}), 
\label{ni}
\ee
where $N_{ij}(t)$ is the matrix of the norm (\ref{norm}) and the rank-one 
matrix $H_{ij}$ equals 
\[
H_{ij}=\frac{1}{(t_i-\xi_1)^2}. 
\]
To find the expectation value one should divide this expression by the 
norm of the eigenstate $|\phi(t)\ra$ which is given by $\det(N_{ij})$.  
Thus the expectation value $\la n_i\ra$ is represented as a ratio of the 
determinants. This formula can be used in the numerical evaluation of 
the occupation number in the case of finite system (for finite N).

 The same method can be used to obtain the determinant representation for 
the two-point correlation function $\la b_i^{+}b_j\ra$. Note that for this 
correlator the final expression was not obtained in ref.\cite{R65}.  
Due to the symmetry of the problem it is sufficient to calculate the average 
$\la\phi(t)|b_2^{+}b_1|\phi(t)\ra$. 
First, we act by the operator $b_1$ on the state $|\phi(t)\ra$ using 
the formula 
\[
b_1\Sigma^{+}(t)=\Sigma^{+}(t)b_1+ \frac{1}{(t-\xi_1)}(1-2n_1). 
\]
Then the action of the operator $b_1$ produce the state 
(see also the Appendix A):  
\[
b_1|\phi(t)\ra=
\sum_i\frac{1}{(t_i-\xi_1)}\left(\Sigma^{+}(t_1)\ldots(i)\ldots\Sigma^{+}(t_M)
\right)\vac 
\]
\[
-2\sum_{i<j}\frac{1}{(t_i-\xi_1)(t_j-\xi_1)}
\left(\Sigma^{+}(t_1)\ldots(i)..(j)\ldots\Sigma^{+}(t_M)\right)b_1^{+}\vac,  
\]
where we denote by $(i)$, $(j)$ the absence of the operators 
$\Sigma^{+}(t_i)$, $\Sigma^{+}(t_i)$ in the last product.  
Clearly, 
the action of the operator $b_2^{+}b_1$ produces the state with two terms 
(see the above formulas and the Appendix A for more details) 
where the first term is equal to  
\[
b_2^{+}b_1|\phi(t)\ra^{(1)} =
\sum_i\frac{1}{(t_i-\xi_1)}\lim_{\zeta\to\xi_2}(\zeta-\xi_2)
\left(\Sigma^{+}(t_1)\ldots\Sigma^{+}(\zeta)\ldots\Sigma^{+}(t_M)\right)\vac,  
\]
where the operator $\Sigma^{+}(\zeta)$ is substituted instead of $i$-th 
operator $\Sigma^{+}(t_i)$. 
Proceeding in the same way as for the average $\la n_i\ra$ we get the similar 
expression for the first term   
\be
\la\phi(t)|b_2^{+}b_1|\phi(t)\ra^{(1)}
=\det(N_{ij}+\tilde{H}^{(1)}_{ij})-\det(N_{ij}), 
\label{ij}
\ee
where the rank-one matrix $\tilde{H}^{(1)}_{ij}$ is equal to 
\[
\tilde{H}^{(1)}_{ij}=\frac{1}{(t_i-\xi_2)^2}\frac{(t_j-\xi_2)}{(t_j-\xi_1)}. 
\]
The second term for $b_2^{+}b_1|\phi(t)\ra$ contains the double limit 
of the form    
\[
-2\sum_{i<j}\frac{1}{(t_i-\xi_1)(t_j-\xi_1)}
\lim_{\zeta_1\to\xi_1}\lim_{\zeta_2\to\xi_2}(\zeta_1-\xi_1)(\zeta_2-\xi_2)
\left(\Sigma^{+}(t_1)...\Sigma^{+}(\zeta_1)...\Sigma^{+}(\zeta_2)...\right)
\vac,  
\]
which means that the two columns should be replaced in the resulting 
determinant. Namely, repeating the above procedure and taking into 
account the form of the matrix $M_{ij}$ (\ref{mrich}) 
we come to the expression 
\[
\la\phi(t)|b_2^{+}b_1|\phi(t)\ra^{(2)}=
-2\sum_{k<l}\frac{1}{(t_k-\xi_1)(t_l-\xi_1)}
\left(\frac{(t_k-\xi_2)(t_l-\xi_1)}{(t_k-t_l)(\xi_2-\xi_1)}\right)
\det_{ij}\left(N_{ij}^{(k,l)}\right), 
\]
where the expression in the parenthesis comes from the factor which can be 
written in front of the determinant in eq.(\ref{mrich}) 
and the matrix $N_{ij}^{(k,l)}$ equals $N_{ij}$ with the exception of 
the two columns $k,l$, which are equal: 
\[
N_{ik}^{(k,l)}=\frac{(t_k-\xi_1)}{(t_i-\xi_1)^2},~~~~
N_{il}^{(k,l)}=\frac{(t_l-\xi_2)}{(t_i-\xi_2)^2}.
\]
In the equivalent form the last expression for the second term is 
\[
\la\phi(t)|b_2^{+}b_1|\phi(t)\ra^{(2)}=
-2\sum_{k<l}\left(\frac{1}{(t_k-t_l)(\xi_2-\xi_1)}\right)
\det_{ij}\left(\tilde{N}_{ij}^{(k,l)}\right), 
\]
where the columns $k,l$ of the new matrix $\tilde{N}_{ij}^{(k,l)}$ are 
\[
\tilde{N}_{ik}^{(k,l)}=\phi_{i}^{(k)}=\frac{(t_k-\xi_2)}{(t_i-\xi_1)^2},~~~~
\tilde{N}_{il}^{(k,l)}=\phi_{i}^{(l)}=\frac{(t_l-\xi_2)}{(t_i-\xi_2)^2}. 
\]
Note that in the particular case of the average $\la b_1^{+}b_1\ra$ due to 
the first term we immidiately get the expression for $\la n_1\ra$, since 
the determinant in the second term is equal to zero due to presence of  
two identical columns in the matrix $\tilde{N}_{ij}^{(k,l)}$.   
Thus, the remarkably simple expression as a sum of determinants is obtained.

Let us perform the similar calculations for the average 
$\la n_{i}n_j\ra$. Considering the action of the operator 
$n_{2}n_1$ on the state $|\phi(t)\ra$, we get the expression similar to 
the second term for $b_2^{+}b_1|\phi(t)\ra$: 
\[
n_{2}n_1|\phi(t)\ra= 
\sum_{i\neq j}\frac{1}{(t_i-\xi_1)(t_j-\xi_2)}
\lim_{\zeta_1\to\xi_1}\lim_{\zeta_2\to\xi_2}(\zeta_1-\xi_1)(\zeta_2-\xi_2)
\left(\Sigma^{+}(t_1)...\Sigma^{+}(\zeta_1)...\Sigma^{+}(\zeta_2)...\right)
\vac. 
\] 
Repeating the calculations performed for the average $\la b_i^{+}b_j\ra$, 
we obtain the following sum of the determinants 
\[
\la\phi(t)|n_{2}n_1|\phi(t)\ra=
\sum_{k\neq l}
\left(\frac{(t_k-\xi_2)(t_l-\xi_1)}{(t_k-t_l)(\xi_2-\xi_1)}\right)
\det_{ij}\left(\tilde{N}_{ij}^{(k,l)}\right), 
\]
where the columns $k,l$ does not depend on the indices $k$ and $l$: 
\[
\tilde{N}_{ik}^{(k,l)}=\phi_{i}^{(k)}=\frac{1}{(t_i-\xi_1)^2},~~~~
\tilde{N}_{il}^{(k,l)}=\phi_{i}^{(l)}=\frac{1}{(t_i-\xi_2)^2}.  
\]
The last expression can also be represented in the equivalent form since,  
taking into account the factors in the parenthesis, 
one can modify the the columns of the matrix $\tilde{N}_{ij}^{(k,l)}$.   
Namely, we have 
\[
\la\phi(t)|n_{2}n_1|\phi(t)\ra=
\sum_{k\neq l}
\frac{1}{(t_k-t_l)(\xi_2-\xi_1)}
\det_{ij}\left(\hat{N}_{ij}^{(k,l)}\right), 
\]
where the two columns of the new matrix $\hat{N}_{ij}^{(k,l)}$ are 
\[
\hat{N}_{ik}^{(k,l)}=\phi_{i}^{(k)}=\frac{(t_k-\xi_2)}{(t_i-\xi_1)^2},~~~~
\hat{N}_{il}^{(k,l)}=\phi_{i}^{(l)}=\frac{(t_l-\xi_1)}{(t_i-\xi_2)^2}.  
\] 
Note that the above expression 
for $\la n_{i}n_j\ra$ is obtained for $i\neq j$. Clearly, at $i=j$ 
the average coincides with $\la n_{i}\ra$. 
Since both averages $\la b_i^{+}b_j\ra$ and $\la n_{i}n_j\ra$ 
are found, the correlator $\la(\sigma_i\sigma_j)\ra$ can also be calculated. 
It is straightforward but lengthy calculation to show that the above 
expressions lead to the expression for $\la(\sigma_i\sigma_j)\ra$ 
found by the variational method and presented in the Appendix B.   
Thus, the remarkably simple determinant expressions for the pair correlators 
of the model suitable for the numerical calculations even at 
the sufficiently large $N$ are obtained.

  As for another physically interesting correlator - the expectation value 
$\la S^{+}S^{-}\ra$, it could be obtained either using the correlator  
$\la b_i^{+}b_j\ra$ obtained above or directly with the help of the 
representation of the average as a certain limit of the scalar product. 
However in fact it is sufficient to calculate the average 
$\la\sum_{i}\xi_{i}n_i\ra$ using the formula (\ref{ni}) which leads to 
the final result 
\be
\la\phi(t)|\sum_{i}\xi_{i}n_i|\phi(t)\ra=
\det(N_{ij}+\hat{H}_{ij})-\det(N_{ij}), 
\label{SS}
\ee
where $\hat{H}_{ij}$ is again the rank-one matrix of the form 
\[
\hat{H}_{ij}=\sum_{\a}\frac{\xi_{\a}}{(t_i-\xi_{\a})^2}. 
\]
Clearly, due to the form of the Hamiltonian (\ref{bcs}), 
from the equation (\ref{SS}) 
one can obtain the expression for the average $\la S^{+}S^{-}\ra$. 
We will show below that, in fact, the sum of eq.(\ref{SS}) and 
the expression for $\la S^{+}S^{-}\ra$ obtained by the different method 
gives the total energy $E$.

Let us note that since the average should be divided by the norm of the 
state in all of the above mentioned cases the expectation values 
of the operators can be represented in the form 
\[
\det_{ij}\left(\d_{ij}+(N^{-1}H)_{ij}\right), 
\]
where $\d_{ij}$- is the Kronecker symbol and the matrix $H_{ij}$ stands for 
one of the three matrices introduced above (\ref{ni}), (\ref{ij}), (\ref{SS}). 
Since in the case of the averages $\la n_i\ra$ and $\la S^{+}S^{-}\ra$ 
(or, equivalently, $\la\sum_{i}\xi_{i}n_i\ra$) the rank-one matrices $H_{ij}$, 
$\hat{H}_{ij}$ depend only on the first index $i$, the last expression can be 
represented in a different form with the help of the identity 
$\det(\d_{ij}+\chi_i)=1+\sum_i\chi_i$. Namely for the first average we get 
\be
\la n_l\ra=\sum_{i,j}\left(N^{-1}\right)_{ij}\phi_j, ~~~~
\phi_i=\frac{1}{(t_i-\xi_l)^2}. 
\label{varni}
\ee
From the equation (\ref{varni}) one readily obtains for the average 
$\la S^{+}S^{-}\ra=\la\sum_{i}\xi_{i}n_i\ra/g-E/g$ 
\be
\la\sum_{i}\xi_{i}n_i\ra=\sum_{i,j}\left(N^{-1}\right)_{ij}\phi_j, ~~~~
\phi_i=\sum_{\a}\frac{\xi_{\a}}{(t_i-\xi_{\a})^2},  
\label{varSS}
\ee  
in agreement with the expression (\ref{SS}) obtained with the help of the 
different method. 
From the equation (\ref{varni}) one can check that the total number of 
particles is $\sum_{i}\la n_i\ra=M$ which can be easily seen from the 
following property of the matrix $N_{ij}$: 
\[
\sum_{\a}N_{i\a}=f(t_i),~~~~~f(t_i)=\sum_{\a}\frac{1}{(t_i-\xi_{\a})^2},   
\]
see eq.(\ref{norm}), or equivalently $1_i=\sum_{j}(N^{-1})_{ij}f(t_j)$.

Now let us turn to the calculation of the average $\la S^{+}S^{-}\ra$ 
in a direct way.  With the help of the formula 
$S^{+}=\lim_{\zeta\to\infty}\zeta\Sigma^{+}(\zeta)$  
we get using the equations (\ref{rich}) that the operator $S^{+}S^{-}$ 
acts on the state $|\phi(t)\ra$ as  
\[
S^{+}S^{-}|\phi(t)\ra=\frac{1}{g}\sum_{i}\lim_{\zeta\to\infty}\zeta 
\left(\Sigma^{+}(t_1)...\Sigma^{+}(\zeta)(i)...\Sigma^{+}(t_M)\right)\vac, 
\]
where $\Sigma^{+}(\zeta)$ stands on the $i$-th place 
(see equation (\ref{a2}) in the Appendix A). Here the factor $1/g$ 
appears due to the equations (\ref{rich}). 
Then taking the limit $\zeta\to\infty$ we finally obtain 
\[
\la\phi(t)|S^{+}S^{-}|\phi(t)\ra=\det(N_{ij}+H_{ij})-\det(N_{ij}), 
\]
with 
\[
   H_{ij}=\frac{1}{g^2},  
\]
where the second factor $1/g$ is due to the last term in the matrix 
(\ref{mrich}). The matrix $H_{ij}$ does not depend on the indices at all. 
Dividing this expression by the norm and repeating the steps described above, 
we obtain the following formula for the average $\la S^{+}S^{-}\ra$: 
\be
\la S^{+}S^{-}\ra= \frac{1}{g^2}\sum_{i,j}\left(N^{-1}\right)_{ij}, 
\label{varSSdir}
\ee
in agreement with the result based on the variational method. 
This form is analogous to the expressions (\ref{varni}), (\ref{varSS}),  
for the other correlators. 
With these three expressions one can check the consistency of the above 
formulas. Namely the sum of (\ref{varSS}) and (\ref{varSSdir}) should 
be equal to the total energy of the system $E$. 
To prove it we use the following equation for the matrix $N_{ij}$: 
\[
\sum_{\a}N_{i\a}t_{\a}=\sum_{\a}\frac{\xi_{\a}}{(t_i-\xi_{\a})^2} 
- \frac{1}{g}, 
\]
which can be easily obtained using the equations (\ref{rich}). Multiplying  
both sides of this equation by the matrix $N^{-1}$, we immediately get the 
energy $E$ as a sum of (\ref{varSS}) and (\ref{varSSdir}).

 In the cases, when the expressions for the correlation 
functions can be obtained with the help of the variation over the parameters 
of the total energy and the integrals of motions \cite{R65}, \cite{G},  
the results coincide with that obtained above as the ratio of 
the determinants. That can be easily proved for the averages $\la n_i\ra$, 
$\la(\s_i\s_j)\ra$, $\la S^{+}S^{-}\ra$ using the formula 
for the matrix inverse to the norm matrix $N_{ij}$ (\ref{norm}). 
We discuss the correspondence of this two approaches in more details 
in the Appendix B.

\vspace{0.3in}

{\bf Conclusion.}

\vspace{0.2in}

In conclusion, for the applications to the realistic fermionic systems,  
it would be interesting to find the realistic discrete - state integrable 
Hamiltonians with the interaction of fermions containing the terms 
describing the breaking of the Cooper pairs. It is possible that 
the models proposed in the present paper can be generalized to this case. 
Let us mention that even in the framework of the BCS model (\ref{bcs}) 
one can include the terms describing the interaction of the single- 
electron states in the following way. Namely, consider the lattice 
Hamiltonian constructed from the fermionic operators 
$c_{i\sigma}^{+}$ ($c_{i\sigma}$) with $\sigma=\uparrow,\downarrow=1,2$: 
\[
H=\sum_i\e_i(n_{1i}+n_{2i})+ 
\sum_{i<j}V_{ij}\left(S_{i}S_{j}\right)- 
g\sum_{i<j}\left(b_{i}^{+}b_{j}+b_{j}^{+}b_{i}\right),  
\]
where $n_{1\sigma}=c_{i\sigma}^{+}c_{i\sigma}$, 
$b_{i}^{+}=c^{+}_{i1}c^{+}_{i2}$ and 
$S_{i}^a={1\over2}(c_i^{+}\sigma_{i}^{a}c_i)$ (the sum over the spin 
indices $1,2$ is implied). 
Since for this Hamiltonian the number of double- occupied sites is 
conserved the eigenstates are given by the superposition of the 
eigenstates of the BCS Hamiltonian and the eigenstates corresponding to 
the second term $\sum_{i<j}V_{ij}(S_{i}S_{j})$.  
Here the coefficients $V_{ij}$ are not necessarily the constant but for 
an integrable model can correspond to any of the integrable quantum spin 
chains (for example, the XXX- spin chain or the Haldane - Shastry spin 
chain in its trigonometric \cite{HS} or hyperbolic \cite{Shyper} versions). 
Note also that using the method presented the Hamiltonians of the 
discrete- state BCS- like models related to the exactly solvable 
$t-J$- models of different symmetry \cite{Sutherland} and the models 
based on the different Lie algebras both in the rational and the 
trigonometric cases can be constructed. 
The main shortcoming of these models from the point of view of 
the applications to the description of electrons close to the Fermi- surface 
is the presence of the single-electron hopping terms. 
In this context the study of Gaudin magnets for the different Lie algebras 
can be useful. 

We have shown that for the calculation of the correlators for the 
Gaudin magnets and the BCS model no special technique for the 
calculation of the scalar products \cite{SKL} is required.  
Using the determinant expressions for the correlation functions 
obtained in the present paper, one could hope to find the 
analytical results for the correlators in the thermodynamic limit 
for the systems with different density of energy levels. 
The determinant expressions for the correlation functions can also be 
useful for the numerical evaluation of correlators both in the case of 
the BCS model with the fixed number of pairs and the general BCS,  
which takes into account the existence of the single - occupied energy levels 
for the excited states. 
In conclusion, let us note, that the similar determinant expressions 
can be obtained both for the BCS and the Gaudin magnets models 
based on the different Lie algebras, which is an interesting problem 
from the theoretical point of view.

\vspace{0.2in}

{\bf Acknowledgments.}

\vspace{0.1in}

This work was supported in part by the RFBI Grant N 00-15-96626.

\vspace{0.2in}

{\bf Appendix A.} 

\vspace{0.1in}

Here we present the most simple and beautiful procedure of 
diagonalization of the Hamiltonian (see for example \cite{G}) 
based on the construction directly in terms of the operators (\ref{sigma}). 
We look for the eigenstate in the form:   
\be
|\phi(t)\ra=\Sigma^{+}(t_1)\Sigma^{+}(t_2)\ldots\Sigma^{+}(t_M)\vac
\label{t}
\ee
We use the following commutational relations for the operator 
$\Sigma^{+}(t)$ which can be easily proved directly or derived from the 
general formulas (\ref{sigma}):  
\[
n_1\Sigma^{+}(t)=\Sigma^{+}(t)n_1+\frac{1}{(t-\xi_1)}b_1^{+},  
\]
\be
b_1\Sigma^{+}(t)=\Sigma^{+}(t)b_1+ \frac{1}{(t-\xi_1)}(1-2n_1). 
\label{com} 
\ee
First, using the first relation, we obtain the formula for the action 
of the operator $n_{\a}$ to the state (\ref{t}) $|\phi(t)\ra=|t\ra$: 
\[
n_{\a}|t\ra=\sum_i\frac{1}{(t_i-\xi_{\a})}
\Sigma^{+}(t_1)\ldots(i)\ldots\Sigma^{+}(t_M)b_{\a}^{+}\vac,
\]
where the sign $(i)$ means the absence of the operator $\Sigma^{+}(t_i)$ 
in the product. Then considering the sum $\sum_{\a}\xi_{\a}n_{\a}$ we obtain 
the following expression for the first part of the Hamiltonian: 
\be
\sum_{\a}\xi_{\a}n_{\a}|t\ra=-S^{+}\sum_i|t(i)\ra+
\left(\sum_{i}t_i\right)|t\ra, 
\label{a1}
\ee
where we denote by the sign $|t(i)\ra$ the state (\ref{t}) without the 
single operator $\Sigma^{+}(t_i)$. 
Next, consider the action of the term $-gS^{+}S^{-}$ on the state (\ref{t}) 
using the second commutational relation (\ref{com}). 
First, considering the action of the operator $b_1$, we obtain the formulas: 
\[
b_1|t\ra=\sum_i\frac{1}{(t_i-\xi_1)}\left( 
\Sigma^{+}(t_1)\ldots(1-2n_i)\ldots\Sigma^{+}(t_M)\right)\vac= 
\]
\[
\sum_i\frac{1}{(t_i-\xi_1)}|t(i)\ra-2\sum_{i<j}\frac{1}{(t_i-\xi_1)(t_j-\xi_1)}
\left(\Sigma^{+}(t_1)\ldots(i)..(j)\ldots\Sigma^{+}(t_M)\right)b_1^{+}\vac. 
\]
After the simple algebraic transformations, using the definition of the 
operator $\Sigma^{+}(t)$, we obtain the following expression:  
\[
S^{-}|t\ra=
\sum_i\left(\sum_{\a}\frac{1}{(t_i-\xi_{\a})}\right)|t(i)\ra+ 
2\sum_{i<j}\frac{1}{(t_i-t_j)}\left(|t(j)\ra-|t(i)\ra\right),  
\]
which finally leads to the result: 
\be
S^{+}S^{-}|t\ra=\sum_i\left(\sum_{\a}\frac{1}{t_i-\xi_{\a}}
-2\sum_{\a\neq i}\frac{1}{t_i-t_{\a}}\right)S^{+}|t(i)\ra. 
\label{a2}
\ee
Thus combining the equations (\ref{a1}) and (\ref{a2}) we get 
\[
H|\phi(t)\ra=E|\phi(t)\ra
-g\sum_i\left(\sum_{\a}\frac{1}{t_i-\xi_{\a}}
-2\sum_{\a\neq i}\frac{1}{t_i-t_{\a}}+\frac{1}{g}\right)S^{+}|t(i)\ra, 
\]
and obtain the eigenvalue $E=\sum_{i}t_i$ and the condition 
of the cancelation of the ``unwanted'' terms $S^{+}|t(i)\ra$ - the equations 
(\ref{rich}): 
\[
\sum_{\a}\frac{1}{t_i-\xi_{\a}}-2\sum_{\a\neq i}\frac{1}{t_i-t_{\a}}=
 -\frac{1}{g}. 
\]
It is straightforward to find also the eigenvalues of the conserved operators 
(\ref{hi}) using this method. In fact, performing the similar calculations, 
we finally obtain the formula for the action of the operators $H_i$ (\ref{hi}) 
to the state $|\phi(t)\ra$ (\ref{t}). For the operator 
\[
H_1=-\frac{1}{g}n_1+\frac{1}{2}\sum_{j\neq 1}
\frac{(\sigma_1\sigma_j)}{(\xi_1-\xi_j)} 
\]
we get the expression 
\[
H_1 |t\ra=\left(\frac{1}{2}\sum_{\a}\frac{1}{(\xi_1-\xi_{\a})}+ 
\sum_{i}\frac{1}{(t_i-\xi_{1})}\right)|t\ra - 
\]
\[
\sum_i\left[\sum_{\a}\frac{1}{t_i-\xi_{\a}}
-2\sum_{k\neq i}\frac{1}{t_i-t_k}+\frac{1}{g}\right] 
\frac{1}{(t_i-\xi_1)}b_1^{+}|t(i)\ra. 
\]
The condition of cancelation of the ``unwanted'' terms $b_1^{+}|t(i)\ra$ 
is equivalent to the equations (\ref{rich}) and the eigenvalues of the 
operators $H_i$ are given exactly by the equation (\ref{ei}),  
however, the last expression for $H_1|t\ra$ is valid for the state 
of the form (\ref{t}) for an arbitrary sets of the parameters 
$\{t\}$ and $\{\xi\}$. 

Note that the similar expression for $H_i|t\ra$ can be easily obtained 
also for the trigonometric case even without the detailed calculations. 
In fact, it is clear, that the analog of the last formula has the form 
\[
H_1|t\ra=E_1(t,\xi)|t\ra+
\sum_{i}f(t_i)\frac{1}{\sin(t_i-\xi_1)}b_{1}^{+}|t(i)\ra, 
\]
where $E_1(t,\xi)$ is the eigenvalue of the operator $H_1$ and the 
function $f(t_i)$ equals 
\[
f(t_i)=\sum_{\a}\ctg(t_i-\xi_{\a})-2\sum_{\a}\ctg(t_i-t_{\a})+1/g, 
\]
so that the trigonometric Richardson's equations have the form 
$f(t_i)=0$. The coefficient in front of the second term in the 
formula for $H_1|t\ra$ can be fixed from the known expression for 
the action of the operator $n_1$ contained in $H_1$.

\vspace{0.2in}

{\bf Appendix B.} 

\vspace{0.1in}

Here we compare the results for the correlators obtained with the help of 
the Algebraic Bethe Ansatz method with the results obtained by means of 
the variation over the parameters \cite{R65}, \cite{G}. Variation over the 
parameters $\xi_i$ and $g$ gives the equation: 
\[
\sum_{\a}^N\frac{\d t_i}{(t_i-\xi_{\a})^2}
-2\sum_{\a\neq i}^M\frac{\d t_i-\d t_{\a}}{(t_i-t_{\a})^2}= 
\sum_{\beta}^N\frac{\d\xi_{\beta}}{(t_i-\xi_{\beta})^2}+\frac{\d g}{g^2}.  
\]
For the case of the averages $\la n_i\ra$, $\la S^{+}S^{-}\ra$  i.e. varying 
over the parameters $\xi_i$, $g$ this equation leads to  
\[
\sum_{j}N_{ij}\d t_j=\sum_{\beta}\phi_{i\beta}\d\xi_{\beta}+\d g/g^2,  
\]
where $N_{ij}$ is the norm matrix (\ref{norm}) and the matrix 
$\phi_{i\beta}=1/(t_i-\xi_{\beta})^2$. 
The solution of this system of equations gives the variation of the energy 
$\d E=\sum_i\d t_i$ which determines the above averages. 
One can easily see that for the correlators $\la n_i\ra$, 
$\la S^{+}S^{-}\ra$ and also $\la\sum_{i}\xi_{i}n_i\ra$ the 
expressions obtained in Section 4 starting from the determinant 
expressions are reproduced. 

In order to calculate the average $\la(\sigma_{i}\sigma_{j})\ra$, 
we consider the variations of the eigenvalues (\ref{ei}) of the 
commuting operators (\ref{hi}) over the parameters $\xi_i$: 
\[
\frac{\d H_i}{\d\xi_j}=\frac{1}{2}\frac{1}{(\xi_i-\xi_j)^2} 
\left(\sigma_{i}\sigma_{j}\right).   
\]
We get for the eigenvalues 
\[
\frac{\d E_i}{\d\xi_j}= \frac{1}{2}\frac{1}{(\xi_i-\xi_j)^2} - 
\sum_{l}\frac{1}{(t_l-\xi_i)^2}\frac{\d t_l}{\d\xi_j}, 
\]
where according to the formulas of Section 4 the matrix 
$\d t_l/\d\xi_{\beta}$ equals: 
\[
\frac{\d t_l}{\d\xi_{\beta}}=\sum_{j,\beta}\left(N^{-1}\right)_{lj}
\frac{1}{(t_j-\xi_{\beta})^2}. 
\]
Using these formulas we easily obtain for the average the expression 
\[
\la\left(\sigma_{i}\sigma_{j}\right)\ra=1-2\left(\xi_i-\xi_j\right)^2 
\sum_{l,k}\frac{1}{(t_l-\xi_i)^2}\left(N^{-1}\right)_{lk}
\frac{1}{(t_k-\xi_j)^2}.  
\]
which can be shown to be in agreement with the determinant expression 
given in ref.\cite{G}. In fact it is straightforward to represent this 
average as a ratio of the determinants: 
\[
\la\left(\sigma_{i}\sigma_{j}\right)\ra=1-2\left(\xi_i-\xi_j\right)^2 
\det(R_{ij})/\det(N_{ij}), 
\]
where the matrix $R_{ij}$ is $(M+1)\times(M+1)$- matrix with the indices 
$i,j=0,1,\ldots M$ and with the matrix elements equal to $R_{00}=0$ and 
\[
R_{kl}=N_{kl},~~~ R_{k0}=\frac{1}{(t_{k}-\xi_j)^2},~~~~ 
R_{0l}=\frac{1}{(t_{l}-\xi_i)^2}
\]
for $k,l=1\ldots M$. Thus the determinant expression is obtained.

\vspace{0.2in}

{\bf Appendix C}

\vspace{0.1in}

Using the formulae obtained above, we present here the solution of 
the modified Knizhnik-Zamolodchikov equations for the vector- valued 
function $|\Phi(\xi)\ra$ of $N$ variables $\xi_1\ldots\xi_N$ 
corresponding to the $SL(2)$ algebra:   
\be
\left(\gamma\frac{\pd}{\pd\xi_i}-\frac{1}{g}n_i+\frac{1}{2}
\sum_{l\neq i}\frac{(\sigma_i\sigma_l)}{(\xi_i-\xi_l)}\right)
|\Phi(\xi)\ra=0. 
\label{kz}
\ee
The modification corresponds to the term $\sim n_i$ and at $g$ 
equal to infinity and the additional parameter $\gamma=1$ 
we obtain the usual Knizhnik-Zamolodchikov equation 
\cite{KZ} for the correlators of the conformal field theory  
(WZW- model). We consider the rational case although the same 
procedure can be performed for the trigonometric case. 
Our presentation follows the 
solution \cite{BF} and make use of the eigenstates of the BCS-  
model (or twisted six-vertex model in the quasiclassical limit). 
However, we present here the solution which is different (in the form)  
from the approach \cite{BF}, and based on the off-shell Bethe Ansatz 
equations for the model (\ref{bcs}), or, equivalently, on the equations 
for the six-vertex model after the quasiclassical limit was 
already performed. For simplicity consider the case $\gamma=1$.   

Following ref.\cite{BF} let us seek for the solution of the equation 
(\ref{kz}) in the form: 
\[
|\Phi(\xi)\ra=\oint dt~\chi(t,\xi)|\phi(t)\ra, 
\]
where $|\phi(t)\ra$ (\ref{t}) is the eigenstate of the Hamiltonian 
(\ref{bcs}),  considered as a vector- valued function of the variables 
$t_i$ and $\xi_j$, and $\chi(t,\xi)$ is some function to be specified below. 
We denote by the symbol $\oint dt$ the integration over the variables 
$t_1\ldots t_M$ in the complex plane over some closed contours 
$C_1\ldots C_M$. The particular solution of the equation depends 
on the choice of the contours.  
 Taking the derivative of this vector- valued function over the 
variable $\xi_1$ gives 
\be
\frac{\pd}{\pd\xi_1}|\Phi(\xi)\ra= 
\oint dt\left(\frac{\pd}{\pd\xi_1}\chi(t,\xi)\right)|\phi(t)\ra+ 
\oint dt~\chi(t,\xi)\frac{\pd}{\pd\xi_1}|\phi(t)\ra. 
\label{deriv}
\ee
Differentiating the eigenstate $|\phi(t)\ra$ we obtain 
\[
\frac{\pd}{\pd\xi_1}|\phi(t)\ra=\sum_i\frac{1}{(t_i-\xi_1)^2}b_1^{+}|t(i)\ra, 
\]
where the state $|t(i)\ra$ is the state (\ref{t}) without the single 
operator $\Sigma^{+}(t_i)$. Since we assume the closed contours of 
integration over $t_i$ and the integration by parts can be applied,  
the second term in the equation (\ref{deriv}) can be represented as 
\[
\sum_{i} \int dt~\left(\frac{\pd}{\pd t_i}\chi(t,\xi)\right) 
\frac{1}{(t_i-\xi_1)}b_1^{+}|t(i)\ra. 
\]
To rewrite this term we make use of the obvious similarity between 
this formula and the expression for $H_1|\phi(t)\ra$, given by the 
last formula in the Appendix A, which is valid for an arbitrary 
parameters $t_i$ and $\xi_{\a}$. In fact, one can represent this 
equation in the equivalent form as
\[
\sum_{i}f(t_i)\frac{1}{(t_i-\xi_1)}b_1^{+}|t(i)\ra=
\left(E_1(t,\xi)-H_1\right)|t\ra, 
\]
where 
\[
f(t_i)= \sum_{\a}\frac{1}{t_i-\xi_{\a}} 
-2\sum_{k\neq i}\frac{1}{t_i-t_k}+\frac{1}{g}.
\]
If one can choose the function $\chi(t,\xi)$ in such a way that 
$(\pd/\pd t_i)\chi(t,\xi)=f(t_i)\chi(t,\xi)$ and the last term takes 
the form 
\[
 \oint dt~\chi(t,\xi)\left(E_1(t,\xi)-H_1\right)|\phi(t)\ra, 
\]
where $E_1(t,\xi)$ is the eigenvalue of the operator $H_1$ (\ref{ei}),  
and the operator $H_1$ is obviously does not depends on $t_i$, 
then we immediately find that the function $|\Phi(\xi)\ra$ is the 
solution of the equation 
\[
\left(\frac{\pd}{\pd\xi_1}+H_1\right)|\Phi(\xi)\ra=0, 
\]
provided the function $\chi(t,\xi)$ satisfies the equation
$(\pd/\pd\xi_1)\chi(t,\xi)=E_{1}(t,\xi)\chi(t,\xi)$. 
Thus we get the following system of the equations for the function 
$\chi(t,\xi)$: 
\[
\frac{\pd}{\pd\xi_i}\chi(t,\xi)=
\left(\sum_j\frac{1}{t_j-\xi_i}+
\frac{1}{2}\sum_{\a\neq i}\frac{1}{\xi_i-\xi_{\a}}\right)\chi(t,\xi),   
\]
\[
\frac{\pd}{\pd t_i}\chi(t,\xi)=\left(\sum_{\a}\frac{1}{t_i-\xi_{\a}} 
-2\sum_{k\neq i}\frac{1}{t_i-t_k}+\frac{1}{g}\right)\chi(t,\xi). 
\]
One can see that this two sets of the equations are compatible and 
the function $\chi(t,\xi)$ is given by 
\[
\chi(t,\xi)=e^{\sum_{i}t_i/g}\prod_{i<j}(t_i-t_j)^{-2}
\prod_{\a<\beta}(\xi_{\a}-\xi_{\beta})^{1/2}\prod_{i,\a}(t_i-\xi_{\a})
\]
Remarkably, the variation of this function over the variables $t_i$ 
gives the equations (\ref{rich}) if one assumes that $t_i$ are at 
the stationary points corresponding to the function $\chi(t,\xi)$  
($(\pd/\pd t_i)\chi(t,\xi)=0$). For the parameter $\gamma\neq 1$ 
the similar expression for the function $\chi(t,\xi)$ can be 
obtained (the powers of the factors in the pruducts should be 
divided by $\gamma$). In the similar way the solution of the 
equations based on the different Lie algebras can be obtained.    

In the trigonometric case the solutions of the $SU(2)$ KZ - 
equations have the similar form: 
\[
\frac{\pd}{\pd\xi_i}\chi(t,\xi)=E_i(t,\xi)\chi(t,\xi),~~~~
\frac{\pd}{\pd t_i} \chi(t,\xi)=f(t_i,t,\xi)\chi(t,\xi),  
\]
where the expressions for $E_i(t,\xi)$ and $f(t_i,t,\xi)$ should be 
substituted by their trigonometric analogs presented above: 
\[
E_i=1/2g-\sum_j\ctg(t_j-\xi_i),~~
f(t_i)=\sum_{\a}\ctg(t_i-\xi_{\a})-2\sum_{\a}\ctg(t_i-t_{\a})+1/g. 
\]
From these equations the explicit solution for the function $\chi(t,\xi)$ 
in the trigonometric case can be obtained. 

Let us mention that in the rational case, except the commuting differential 
operators $\pd/\pd\xi_i-H_i$, in the $SU(2)$ case there is an extra 
commuting differential operator $g^2\pd/\pd g-H$, where $H$ is the 
Richardson Hamiltonian (\ref{bcs}), (\ref{Richard}) and $g$ is the 
corresponding coupling constant: 
\[
\left[ g^2\frac{\pd}{\pd g}-H~;~\frac{\pd}{\pd\xi_i}-H_i\right]=0 
\]
(for example, see \cite{Resh}, \cite{BK}). 
In conclusion, let us mention that the integration contours 
$C_1,\ldots C_M$, which define the solution of the modified KZ - equations, 
are not necessarily the closed contours, which do not intersect the 
branch cuts of the integrand in the complex plane. 
The only condition is that the integral over the total derivative of the 
form 
\[
\sum_i\oint dt \frac{\pd}{\pd t_i}\left(\chi(t,\xi)
\frac{1}{(t_i-\xi_{\a})}b_{\a}^{+}|t(i)\ra\right)=0 
\]
for each $\a=1,\ldots N$, which follows from the derivation presented above.

\vspace{0.2in}

{\bf Appendix D} 

\vspace{0.1in}

  Here we present the analytical solution of the BCS model in the 
continuum limit (i.e. in the limit $N\to\infty$) for the case of the 
equal- spacing distribution of the energy levels $\xi_i$, or for the 
density of energy levels $\rho(\xi)$ equal to unity at some interval 
which can be chosen as $\xi\in(-1,1)$. Although the solution of the 
equations (\ref{rich}) for the continuum limit has been considered 
previously \cite{G}, \cite{R77}, with the help of the electrostatic 
analogy, and the final solution in agreement with the result of the 
variational BCS treatment (\ref{bcs}) was obtained, some questions 
remained obscure. For instance, the choise of the ansatz for the 
complex electric field with a branch cut along the line is not well 
understood. Thus the additional arguments and approachs for this 
problem are highly desirable. 
First of all, note that the equations (\ref{rich}) 
can be represented as the conditions of the minimum of the ``energy'' 
functional $\pd/\pd t_i\Phi(t)=0$, where the roots $t_i$ are 
considered as a positions of charges at the two- dimensional complex 
plane interacting through the Coulomb potential and subjected to the 
homogeneous electric field of the strength $-1/g$, directed along the 
real axis,  
\[
\Phi(t)=\sum_{i,\a}\ln|t_i-\xi_{\a}|-2\sum_{i<j}\ln|t_i-t_j|+ 
\frac{1}{g}\sum_{i}\Re t_i. 
\]
The potential between the like charges $t_i$ of the value $+1$ 
corresponds to the repulsion, while the their interaction with 
the points $\xi_i$ with the charges $-1/2$ is repulsive.    

 Let us denote by $h(z)$ the holomorphic (which depends 
only on the coordinate $z$ at the part of the complex plane without 
the charges) complex electric field defined in such a way that the 
integral of $h(z)$ over some closed contour $C$ in the comlex plane 
is equal $\oint_{C}dzh(z)=2\pi iq$, where $q$ is the total electric 
charge enclosed by the contour $C$. Clearly, the unit charge, centered 
at the origin, gives the complex electric field equal to $1/z$. 

First, let us suppose that in the continuum limit the roots $t_i$ 
form some curve $\Gamma$, symmetric with respect to the real axis,  
with the continuous complex density of roots along the curve $R(t)$ 
(we assume here that all roots $t_i$ appears in a complex pairs).  
Let us denote by $a$ and $b=a^{\star}$ the endpoints of the curve 
$\Gamma$, and by $C$- the contour enclosing the curve $\Gamma$.  
We also assume that the curve $\Gamma$ does not intersect the 
domain of the charges $\xi$ at the real axis. 
  The discontinuity of $h(z)$ at $\Gamma$ is equal to the density 
$R(t)$, $\Delta h(z)=2\pi iR(z)$, which by the Cauchy theorem means that 
$h(z)$ is the analytic continuation of $R(z)$ from the curve $\Gamma$. 
In particular that means that the total number of particles is
$\oint_{C}dz~h(z)=2\pi iM$ and the total energy is 
$\oint_{C}dz~zh(z)=2\pi iE$ 

The Gaudin's assumption is that the field $h(z)$ has the branch cut 
along the line $(a,b)$ of the form 
\[
h(z)=\sqrt{(z-a)(z-b)}\int\frac{d\xi}{\xi-z}\phi(\xi),
\]
where the function $\phi(\xi)$ can be fixed from the condition that 
the residues of the field at the points $\xi$ should be equal to $-1/2$: 
\[
\phi(\xi)= (1/2) \left((\xi-a)(\xi-b)\right)^{-1/2}.  
\]
Since at infinity $|z|\to\infty$ the field should be equal to $-1/g$, 
we have the equation 
\[
\int d\xi\phi(\xi)=\frac{1}{2g}.   
\]
The conservation of the total number of particles gives the 
equation $\int_{\Gamma}dtR(t)=M$, and the total energy equals   
$E=\int_{\Gamma}dttR(t)$.

Alternatively, one can write down the equations (\ref{rich}) in the 
continuum limit in the following form: 
\[
\int d\xi\frac{1}{t-\xi}-2\int_{\Gamma}dz\frac{R(z)}{t-z}=-\frac{1}{g}, 
\]
where the integal over the curve $\Gamma$ can be transformed to the 
integral over the contour $C$ as 
\[
\int d\xi\frac{1}{t-\xi}-2\oint_{C}\frac{dz}{2\pi i}h(z)\frac{1}{t-z}
=-\frac{1}{g}. 
\]  
Substituting the ansatz for $h(z)$, we obtain 
\[
\int d\xi\frac{1}{t-\xi}-
2\int d\xi\phi(\xi)\oint_{C}\frac{dz}{2\pi i} 
\frac{\sqrt{(z-a)(z-b)}}{(t-z)(\xi-z)}=-\frac{1}{g}, 
\]
and transforming the integration contour $C$ into the two contours, 
the first one is around the domain of $\xi$, 
and the second one is at the circle at the infinity, we get 
\[
\int d\xi\frac{1}{t-\xi}-2\left[ 
\int\frac{d\xi}{t-\xi}\phi(\xi)\sqrt{(\xi-a)(\xi-b)}+\int d\xi\phi(\xi) 
\right] =-\frac{1}{g}. 
\]
This equation leads to the two equations, the first one is the 
equation for the function $\phi(\xi)$ and the second one is for the 
integral $\int d\xi\phi(\xi)=1/2g$.  Applying the same transformation 
of the integration contour to the integrals $M=\oint_{C}(dz/2\pi i)h(z)$ 
and $E=\oint_{C}(dz/2\pi i)zh(z)$, and substituting the value 
$a(b)=\pm i\Delta$, we obtain the gap equation, the equation for the 
particle number and the energy in agreement with the BCS theory. 
To invetsigate the $1/N$ corrections Richardson \cite{R77} has derived 
the closed Riccati type integro-differential equation for the electric 
field 
\[
h(z)=\sum_i\frac{1}{z-t_i}-\frac{1}{2}\sum_{\a}\frac{1}{z-\xi_{\a}} 
-\frac{1}{g} 
\]
in the continuum limit, using the same operation with the contour 
integration in the complex plane as above. However, at present time, the 
solution of this equation (without using the ansatz for the field $h(z)$)  
is absent. The form of the curve $\Gamma$ can be obtained from 
the condition that the component of the electric field along the curve 
should be equal to zero for the point at the curve $\Gamma$.  
One can imagine the curve $\Gamma$ as a metallic plate of the special 
form with the endpoints $a$, $b$. At each point near this plate the 
vector of the electric field has a direction perpendicular to the plate. 
In particular that means that the electic field at the points $a$, $b$ is 
equal to zero, which is fulfilled for the ansatz for $h(z)$.  
In other words, the curve $\Gamma$ can be found from the condition that 
it should be the equipotential curve for the electric field $h(z)$. 
One should stress that the form of the branch cut between the points 
$a$ and $b$ can be chosen in an arbitrary way, and for the ansatz 
for $h(z)$ it is assumed that the branch cut coincides with the 
curve $\Gamma$. 
Note that one can calculate the density of charges $|R(t)|$ along the 
curve and obtain the form $\sim((t-a)(t-b))^{1/2}f(t)$, where $f(t)$ 
is some smooth function, characteristic for the matrix models. 
Since one can imagine the conformal mapping of $\Gamma$ onto the 
line $(a,b)$, which reduce the problem to the solution of the matrix 
model with some potential, which should exhibit the density of states 
of the same form, this can be considered as a justification 
of the ansatz.

\end{document}